# Constructive Algorithms for Discrepancy Minimization

Nikhil Bansal [*]

**Abstract**

Given a set system $(V, \mathcal{S})$, $V = \{1, \ldots, n\}$ and $\mathcal{S} = \{S_1, \ldots, S_m\}$, the minimum discrepancy problem is to find a 2-coloring $\mathcal{X} : V \to \{-1, +1\}$, such that each set is colored as evenly as possible, i.e. find $\mathcal{X}$ to minimize $\max_{j \in [m]} \left| \sum_{i \in S_j} \mathcal{X}(i) \right|$.

In this paper we give the first polynomial time algorithms for discrepancy minimization that achieve bounds similar to those known existentially using the so-called Entropy Method. We also give a first approximation-like result for discrepancy. Specifically we give efficient randomized algorithms to:

1. Construct an $O(n^{1/2})$ discrepancy coloring for general sets systems when $m = O(n)$, matching the celebrated result of Spencer [17] up to constant factors. Previously, no algorithmic guarantee better than the random coloring bound, i.e. $O((n \log n)^{1/2})$, was known. More generally, for $m \geq n$, we obtain a discrepancy bound of $O(n^{1/2} \log(2m/n))$.

2. Construct a coloring with discrepancy $O(t^{1/2} \log n)$, if each element lies in at most $t$ sets. This matches the (non-constructive) result of Srinivasan [19].

3. Construct a coloring with discrepancy $O(\lambda \log(nm))$, where $\lambda$ is the hereditary discrepancy of the set system.

The main idea in our algorithms is to produce a coloring over time by letting the color of the elements perform a random walk (with tiny increments) starting from 0 until they reach $-1$ or $+1$. At each time step the random hops for various elements are correlated using the solution to a semidefinite program, where this program is determined by the current state and the entropy method.

## 1 Introduction

Let $(V, \mathcal{S})$ be a set-system, where $V = \{1, \ldots, n\}$ are the elements and $\mathcal{S} = \{S_1, \ldots, S_m\}$ is a collection of subsets of $V$. Given a $\{-1, +1\}$ coloring $\mathcal{X}$ of elements in $V$, let $\mathcal{X}(S_j) = \sum_{i \in S_j} \mathcal{X}(i)$ denote the discrepancy of $\mathcal{X}$ for set $S$. The discrepancy of the collection $\mathcal{S}$ is defined as

$$\mathrm{disc}(\mathcal{S}) = \min_{\mathcal{X}} \max_{j \in [m]} |\mathcal{X}(S_j)|.$$

Understanding the discrepancy of various set-systems has been a major area of research both in mathematics and computer science, and this study has revealed fascinating connections to various areas of mathematics. Discrepancy also has a range of applications to several topics in computer science such as probabilistic and approximation algorithms, computational geometry, numerical integration, derandomization, communication complexity, machine learning, optimization and so on. We shall not attempt to describe these connections and applications here, but refer the reader to [6, 9, 12].

[*]IBM T. J. Watson Research Center, Yorktown Heights, NY 10598. E-mail: *nikhil@us.ibm.com*

## 1.1 Discrepancy of General Set Systems

*What is the discrepancy of an arbitrary set system with $n$ elements and $m$ sets?*
This is perhaps the most basic question in discrepancy theory. Clearly, if we color the elements randomly, for any set $S$, we expect $|\mathcal{X}(S)|$ to be about $O(|S|^{1/2}) = O(n^{1/2})$, i.e. about the standard deviation. Moreover, by standard tail bounds, the probability that $|\mathcal{X}(S)| \geq cn^{1/2}$ is at most $e^{-\Omega(c^2)}$. So, by union bound over the $m$ sets, the discrepancy of the set system will be $O((n \log m)^{1/2})$. This bound for randomly colorings is also tight in general.

Surprisingly, it turns out that better colorings always exist! A celebrated result of Spencer [17] states that: Any set system on $n$ elements and $m \geq n$ sets has $O((n \log(2m/n))^{1/2})$ discrepancy. This guarantee is most interesting when $m = O(n)$. In particular when $m = n$, Spencer showed a bound of $6n^{1/2}$ (commonly referred to as the "six standard deviations suffice" result). This is the best possible bound up to constant factors. Spencer's result is one of the highlights of discrepancy theory and is based on a clever use of the Pigeonhole Principle, a technique first developed by Beck [4]. The technique has since been used widely and is referred to as the Entropy Method or the Partial Coloring Lemma (we discuss this method and its application to obtain Spencer's result in section 2).

However, prior to our work, it was not known how to make this result algorithmic. In fact, no better efficient algorithm than simply random coloring was known and reducing this gap has been a long-standing question [12, 17, 1, 19]. Due to its fundamental use of the Pigeon Hole Principle, Spencer's result is widely believed to be more non-constructive than other existential results such as those based on the probabilistic method or the Lovasz Local Lemma. We quote

*"Is there a polynomial time algorithm that gives discrepancy $Kn^{1/2}$ . . .. The difficulties in converting these theorems to algorithms go back to the basic theorem of this Lecture and lie, I feel, in the use of the Pigeonhole Principle . . . ".* – Joel Spencer [18] (Page 69).

It is also known that any non-adaptive or online algorithm (for details see [2], page 239) must have a discrepancy of $\Omega(\sqrt{n \log n})$, and it has been conjectured [2], page 240, that no polynomial time algorithm may exist for finding a coloring with discrepancy $c\sqrt{n}$.

In this paper we resolve this question and show that.

**Theorem 1.1.** *Given any set system with $n$ elements and $n$ sets $S_1, \ldots, S_n$, there is a randomized polynomial time algorithm that with probability at least $1/\log n$, constructs a $\{-1, +1\}$ coloring $\mathcal{X}$ with discrepancy $O(n^{1/2})$. More generally for $m \geq n$, our algorithm achieves a bound of $O(n^{1/2} \log(2m/n))$ and succeeds with probability at least $1/\log m$.*

We note that for general $m \geq n$, our algorithm has a somewhat worse dependence on $(m/n)$ than the tight $O(n^{1/2} \log(2m/n)^{1/2})$ bound achievable non-constructively. Also, it suffices to consider the case of $m \geq n$: if $m \leq n$, one can essentially reduce $n$ to $m$ using standard techniques [17], implying a (tight) discrepancy of $O(\sqrt{m})$.

## 1.2 Bounded Degree Sets: The Beck-Fiala Setting

Another significant result in discrepancy theory is a theorem due to Beck and Fiala [5]: The discrepancy of any set system $(V, \mathcal{S})$ is at most $2t - 1$, where $t$ is the maximum degree of $(V, \mathcal{S})$, i.e. the maximum number of times an element appears in sets in $\mathcal{S}$.

The proof of this result is algorithmic. This bound was improved slightly to $2t-3$ by Bednarchak and Helm [7], and this is currently best known bound independent of $n$. Beck and Fiala [5] conjectured that the minimum discrepancy is always $O(t^{1/2})$, and this remains a major open question. If the guarantee is allowed to depend on $n$, Beck and Spencer [4, 18] showed that the discrepancy is $O(t^{1/2} \log t \log n)$. Refining their analysis, the bound was improved to $O(t^{1/2} \log n)$ by Srinivasan [19]. Both these proofs are based on the entropy method and are non-constructive. The best known result along these lines is due to Banaszczyk [3] that achieves a bound of $O(t^{1/2} \log^{1/2} n)$. This result is based on certain inequalities for Gaussian measures on $n$-dimensional convex bodies due to [10] and also seems to be inherently existential to the best of our knowledge.

In this paper we give a constructive version of Srinivasan's result.

**Theorem 1.2.** *Given any set system $(V, \mathcal{S})$ with $n$ elements and degree at most t, there is a randomized polynomial time algorithm that with probability at least $1/n$, constructs a $\{-1, +1\}$ coloring $\mathcal{X}$ with discrepancy $O(t^{1/2} \log n)$.*

## 1.3 Pseudo-Approximation and Hereditary Discrepancy

A natural question thus is whether the discrepancy of a particular instance can be approximated efficiently. Very recently Charikar et al.[8] have shown very strong lower bounds for this problem. In particular, they show that there exists set systems with $m = O(n)$ sets, such that no polynomial time algorithm can distinguish whether the discrepancy is 0 or $\Omega(\sqrt{n})$, unless $P = NP$.

Here we prove the following pseudo-approximation result with respect to hereditary discrepancy. Recall that the hereditary discrepancy of a set system $(V, S)$ is defined as the maximum value of discrepancy over all subsets $W$ of $V$. Specifically, given $W \subseteq V$, let $\mathcal{S}_{|W}$ denote the collection $\{S \cap W : S \in \mathcal{S}\}$. Then, the hereditary discrepancy of $(V, \mathcal{S})$ is defined as

$$\mathrm{herdisc}(\mathcal{S}) = \max_{W \subseteq V} \mathrm{disc}(\mathcal{S}_{|W}).$$

We show the following result:

**Theorem 1.3.** *Given any set system $(V, \mathcal{S})$ with hereditary discrepancy at most $\lambda$, there is a randomized polynomial time algorithm that with probability at least $1/n$, constructs a $\{-1, +1\}$ coloring $\mathcal{X}$ with discrepancy $O(\lambda \log(mn))$.*

This answers a question of Matousek [14].

A consequence of our proof of theorem 1.3 is the following: Let us define the hereditary vector discrepancy of a set system $\mathcal{S}$, denoted $\mathrm{hervecdisc}(\mathcal{S})$, as the smallest value of $\lambda$ such that for each subset $W \subseteq V$, the following semi-definite program is feasible.

$$||\sum_{i \in S_j \cap W} v_i||_2^2 \leq \lambda^2 \qquad \text{for each set } S_j \tag{1}$$

$$||v_i||_2^2 = 1 \qquad \forall i \in W \tag{2}$$

Being a relaxation, clearly $\mathrm{hervecdisc}(\mathcal{S}) \leq \mathrm{herdisc}(\mathcal{S})$. Our proof of theorem 1.3 actually produces a coloring with discrepancy $O(\mathrm{hervecdic}(S) \cdot \log(mn))$. Applying theorem 1.3 to each restriction $\mathcal{S}_{|W}$ for $W \subseteq V$ also implies that $\mathrm{herdisc}(\mathcal{S}) = O(\mathrm{hervecdisc}(\mathcal{S}) \cdot \log(mn))$. While do not know how to compute or even approximate $\mathrm{hervecdisc}(\mathcal{S})$ in polynomial time, it might be an interesting quantity to investigate, as any $\beta$ approximation for it would imply an $O(\beta \log(mn))$ approximation for hereditary discrepancy.

## 1.4 Organization

Our algorithms are based on an iterative application of semi-definite programming. In particular, we construct the coloring over time by solving a sequence of semi-definite programs, and use the solution of the SDP to define correlated random walks with tiny increments for each color. The walk for each element continues until it reach $-1$ or $+1$. Interestingly, the non-constructive entropy method is a major component in our algorithm: The semi-definite programs that we construct at each stage are guided by the parameters given by the entropy method.

We give a high-level overview of our method in section 3. We begin in section 2 by describing some preliminary concepts that we need. At the end of section 2, we also describe the entropy method, and show how it is applied to obtain the results of [17] and [19]. In section 4 we prove theorem 1.3 which is technically the simplest result. The ideas developed there also imply theorem 1.2 which is proved in section 4.3. Section 4 lays the basic groundwork for section 5 where we eventually prove theorem 1.1.

## 2 Preliminaries

### 2.1 Gaussian Random Variables

We recall the following standard facts about Gaussian distributions. The Gaussian distribution $N(\mu, \sigma^2)$ with mean $\mu$ and variance $\sigma^2$ has probability distribution function

$$f(x) = \frac{1}{(2\pi)^{1/2}\sigma} e^{-(x-\mu)^2/2\sigma^2}.$$

*Additivity:* If $g_1 \sim N(\mu_1, \sigma_1^2)$ and $g_2 \sim N(\mu_2, \sigma_2^2)$ are independent Gaussian random variables, then for any $t_1, t_2 \in \mathbb{R}$, the random variable

$$t_1 g_1 + t_2 g_2 \sim N(t_1\mu_1 + t_2\mu_2, t_1^2\sigma_1^2 + t_2^2\sigma_2^2).$$

The additivity property of Gaussians implies that

**Lemma 2.1.** *Let $g \in \mathbb{R}^n$ be a random Gaussian, i.e. each coordinate is chosen independently according to distribution $N(0, 1)$. Then for any vector $v \in \mathbb{R}^n$, the random variable $\langle g, v \rangle \sim N(0, ||v||_2^2)$. Here as usual, $||v||_2 = (\sum_i v(i)^2)^{1/2}$ denotes the $\ell_2$ norm of $v$.*

### 2.2 Probabilistic Tail Bounds for Martingales

We will use the following probabilistic tail bound repeatedly.

**Lemma 2.2.** *Let $0 = X_0 = X_1, \ldots, X_n$ be a martingale with increments $Y_i = X_i - X_{i-1}$. Suppose for $1 \leq i \leq n$, we have that $Y_i | (X_{i-1}, \ldots, X_0)$ is distributed as $\eta_i G$, where $G$ is a standard Gaussian $N(0, 1)$ and $\eta_i$ is a constant such that $|\eta_i| \leq 1$ (note that $\eta_i$ may depend on $X_0, \ldots, X_{i-1}$). Then,*

$$\Pr[|X_n| \geq \lambda\sqrt{n}] \leq 2e^{-\lambda^2/2}.$$

*Proof.* Let $\alpha$ be a parameter to be optimized later. We have,

$$\begin{aligned}
\mathbb{E}[e^{\alpha Y_i} | X_{i-1}, \ldots, X_0] &\leq \int_{-\infty}^{\infty} e^{\alpha y} \cdot \left(\frac{1}{(2\pi)^{1/2}\eta_i} e^{-y^2/2\eta_i^2}\right) dy \\
&= e^{\alpha^2 \eta_i^2/2} \cdot \int_{-\infty}^{\infty} \left(\frac{1}{(2\pi)^{1/2}\eta_i} e^{-(y-\alpha\eta_i^2)^2/2\eta_i^2}\right) dy \\
&= e^{\alpha^2 \eta_i^2/2} \leq e^{\alpha^2/2}.
\end{aligned}$$

Now,

$$\mathbb{E}[e^{\alpha X_n}] = \mathbb{E}[e^{\alpha X_{n-1}} e^{\alpha Y_n}] = \mathbb{E}[e^{\alpha X_{n-1}} \mathbb{E}[e^{\alpha Y_n} | X_{n-1}, \ldots, X_0]] \leq e^{\alpha^2/2} \mathbb{E}[e^{\alpha X_{n-1}}].$$

Thus it follows by induction that $\mathbb{E}[e^{\alpha X_n}] \leq e^{\alpha^2 n/2}$. Finally,

$$\Pr[X_n \geq \lambda\sqrt{n}] = \Pr[e^{\alpha X_n} \geq e^{\alpha\lambda\sqrt{n}}] \leq e^{-\alpha\lambda\sqrt{n}}\mathbb{E}[e^{\alpha X_n}] \leq e^{-\alpha\lambda\sqrt{n}+\alpha^2 n/2}.$$

Setting $\alpha = \lambda/\sqrt{n}$ and noting that $\Pr[X_n \geq \lambda\sqrt{n}] = \Pr[X_n \leq -\lambda\sqrt{n}]$ implies the claim. □

## 2.3 Semidefinite Programming

Let $M_n$ denote the class of all symmetric $n \times n$ matrices with real entries. For two matrices $A, B \in \mathbb{R}^{n \times n}$, the Frobenius inner product of $A$ and $B$ is defined as $A \bullet B = \operatorname{tr}(A^T B) = \sum_{i=1}^n \sum_{j=1}^n a_{ij} b_{ij}$. For $Y \in \mathbb{R}^{n \times n}$, let $Y \succeq 0$ denote that it is semidefinite, i.e. all its eigenvalues are non-negative. Then a general semidefinite program has the following form

$$\begin{aligned}
\max \quad & C \bullet Y \\
\text{s.t.} \quad & D_i \bullet Y \leq d_i, \quad 1 \leq i \leq k \\
& Y \succeq 0 \\
& Y \in M_n
\end{aligned}$$

where $C, D_1, \ldots, D_k \in M_n$ and $d_1, \ldots, d_k$ are real numbers.

Semidefinite programs form an important class of convex programs and can be solved efficiently to any desired level of accuracy. Since $Y$ is a symmetric semidefinite matrix, it can be written as $Y = W^T W$ for some $W \in \mathbb{R}^n$. Let $y_{ij}$ denote the $(i,j)$-entry of $Y$ and let $w_i$ be the $i$-th column of $W$, then $y_{ij} = \langle w_i, w_j \rangle$ for each $i, j$. Thus, one can equivalently view an SDP as an arbitrary linear program on variables of the form $\langle w_i, w_j \rangle$ where $w_i \in R^m$ for some $m$ (however, in the SDP solution, one cannot control the dimension $m$ of the vectors $w_i$. In general $m$ could be as high as the number of vectors $w_i$). We refer the reader to [20] for further details about semidefinite programming.

## 2.4 The Entropy Method

We recall here the partial coloring lemma of Beck [4], based on the Entropy Method. We also describe how it is used to obtain the results of [17] and [19]. The form we present below is from [13].

**Lemma 2.3** (Entropy Method). *Let $\mathcal{S}$ be a set system on an $n$-point set $V$, and let a number $\Delta_S > 0$ be given for each set $S \in \mathcal{S}$. Suppose $\Delta_S$ satisfy the condition*

$$\sum_{S \in \mathcal{S}} g\left(\frac{\Delta_S}{\sqrt{|S|}}\right) \leq \frac{n}{5} \tag{3}$$

*where*

$$g(\lambda) = \begin{cases} Ke^{-\lambda^2/9} & \text{if } \lambda > 0.1 \\ K\ln(\lambda^{-1}) & \text{if } \lambda \leq 0.1 \end{cases}$$

*and $K$ is some absolute constant (wlog we will assume that $K > 3$). Then there is a partial coloring $\mathcal{X}$ that assigns $-1$ or $+1$ to at least $n/2$ variables (and $0$ to the rest of the variables), and satisfies $|\mathcal{X}(S)| \leq \Delta_S$ for each $S \in \mathcal{S}$.*

This result is proved by arguing (via an entropy/counting argument) that there are exponentially many colorings $\mathcal{X}_1, \ldots, \mathcal{X}_\ell$ such that for every $i, j$, $1 \leq i < j \leq \ell$, the difference in discrepancy $|\mathcal{X}_i(S) - \mathcal{X}_j(S)| \leq \Delta_S$ for all $S$. Since $\ell$ is exponential, there must exist two colorings among these $\ell$, say $\mathcal{X}_1$ and $\mathcal{X}_2$, that differ on $\Omega(n)$ coordinates. Then, $(\mathcal{X}_1 - \mathcal{X}_2)/2$ gives the desired partial coloring.

**Spencer's Result [17]:** The coloring is constructed in phases. In phase $i$, for $i = 0, \ldots, \log n$, the number of uncolored elements is at most $n_i \leq n/2^i$. In phase $i$, apply lemma 2.3 to these $n_i$ elements with $\Delta_S^i = c(n_i \log(2m/n_i))^{1/2}$. It is easily verified that (3) holds for a large enough constant $c$. This gives a partial coloring on at least $n_i/2$ elements, with discrepancy for any set $S$ at most $\Delta_S^i$. Summing up over the phases, the overall discrepancy for any set is at most

$$\Delta_S^i = \sum_i c \left( n 2^{-i} \log \left( \frac{2m}{n 2^{-i}} \right) \right)^{1/2} = O((n \log(2m/n))^{1/2}).$$

**Srinivasan's result [19]:** Again the coloring is constructed in phases $i = 0, \ldots, \log n$, where at most $n_i \leq n/2^i$ elements are uncolored in phase $i$. In phase $i$, let $s_{i,j}$ denote the number of sets with number of uncolored elements in $[2^j, 2^{j+1})$. As the degree of the set system is at most $t$, we have $s_{i,j} \leq \min(m, n_i t/2^j)$. Using this fact, a (careful) calculation shows that (3) can be satisfied if we set $\Delta_S = ct^{1/2}$ for some large enough constant $c$. The $\log n$ phases imply a total discrepancy of $O(t^{1/2} \log n)$.

## 3 Our Approach

We consider a linear variant of colorings, where a coloring is a vector $x \in [-1, 1]^n$ instead of $\{-1, +1\}^n$. Our algorithm constructs the final coloring iteratively in several steps. Let $x_t \in \mathbb{R}^n$ denote the coloring at time $t$. We start with the coloring $x_0 = (0, 0, \ldots, 0)$ initially. We update the coloring over time as $x_t = x_{t-1} + \gamma_t$ by applying suitably chosen (tiny) updates $\gamma_t \in \mathbb{R}^n$. Thus the color $x_t(i)$ of each element $i \in [n]$ evolves over time, until it reaches $-1$ or $+1$. At that time the color of $i$ is considered *fixed* and is never updated again. The procedure continues until all the elements are colored either $-1$ or $+1$.

The updates $\gamma_t$ are chosen carefully (by rounding a certain SDP) and are related to the parameters in the partial coloring lemma as follows: Consider the floating elements at time $t$, i.e. whose color has not been fixed thus far until time $t - 1$. For ease of discussion here, let us assume that all the $n$ elements are floating. Suppose we know the existence (using entropy method or otherwise) of a partial coloring $\mathcal{X}$ on these floating elements, such that $|\mathcal{X}(S)| \leq \Delta_S$ for each $S \in \mathcal{S}$. Then we find a collection of real numbers $\eta_t(i)$, for $i \in [n]$ that satisfy the following properties.

1. *Unbiased Gaussian:* Conditioned upon the evolution of the algorithm until time $t - 1$, each entry $\eta_t(i)$ is distributed as an unbiased Gaussian with standard deviation at most 1.

2. *Large Progress:* The sum of standard deviations of $\eta_t(i)$ over $i \in [n]$ is at least $n/2$.

3. *Low Discrepancy:* The entries $\eta_t(i)$ are correlated such that for every set $S_j$, conditional on the evolution of the algorithm until $t - 1$, the sum $\sum_{i \in S_j} \eta_t(i)$ is distributed as an unbiased Gaussian with standard deviation at most $\Delta_S$.

Then we set $\gamma_t(i) = \gamma \cdot \eta_t(i)$, where $\gamma$ is a small scaling parameter, say for example $\gamma = 1/n$, and update $x_t(i) = x_{t-1}(i) + \gamma_t(i)$ for all $i \in [n]$. By property 1, note the color $x_t(i)$ of each element $i$ forms a martingale, that stops upon reaching $-1$ or $+1$. By properties 1 and 2, at each time step, at least $\Omega(n)$ elements have an increment of magnitude $\Omega(\gamma)$. So after about $O(1/\gamma^2)$ steps, in expectation, about $\Omega(n)$ elements will reach $-1$ or $+1$ and get fixed. Moreover, by property 3, the discrepancy of each set $S$ also forms a martingale with increments of magnitude roughly $O(\gamma \Delta_S)$. Thus in $O(1/\gamma^2)$ steps, the expected discrepancy of set $S$ will be about $O(\Delta_S)$. Note that this gives a procedure that roughly corresponds to the partial coloring lemma: In particular, given any coloring $x \in [-1, 1]^n$ with $a$ floating variables, it produces another coloring (in $O(1/\gamma^2)$ steps) with at most $a/2$ floating variables, such that each set $S$ incurs an additional discrepancy of $\Delta_S$ in expectation.

This already suffices to show theorems 1.3 and 1.2. Let us consider theorem 1.3. We apply the above procedure for $O((\log n)/\gamma^2)$ time steps, until all the variables are fixed to $\{-1,+1\}$. As the hereditary discrepancy is $\lambda$, we can always set $\Delta_S = \lambda$, irrespective of the elements fixed to $\{-1,+1\}$ thus far. This implies an expected discrepancy of $O(\lambda\sqrt{\log n})$ for each set $S$. By standard tail bounds and taking union over the $m$ sets, this implies an $O(\lambda \log(mn))$ discrepancy coloring.

However the above idea by itself does not suffice for theorem 1.1. The problem is that here we want to guarantee that the discrepancy for *every* set is $O(n^{1/2})$, whereas the above idea only gives us discrepancy $O(n^{1/2})$ in *expectation*. So would end up losing a $O(\log^{1/2} n)$ factor due to the union bound over the sets (obtaining nothing better than a random coloring). So, our second idea is to observe that we can control the parameters $\Delta_S$ for each set. We refine the probabilistic procedure above by finely adjusting the parameter $\Delta_S$ for each set $S$ over time, depending on how "dangerous" $S$ has become, while ensuring that $\Delta_S$'s still satisfy the entropy condition (3). To illustrate the idea, we sketch below a simpler $O((n \log \log \log n)^{1/2})$ constructive bound.

Consider the following: Initially, we set all $\Delta_S = cn^{1/2}$ for large enough $c$ so that (3) is satisfied easily and has some slack. As previously, we obtain a corresponding vector $\gamma_t$ and add it to the coloring thus far. We repeat this for $O(1/\gamma^2)$ steps, at which point we expect half the colors to reach either $-1$ or $+1$. During these steps, if the discrepancy $|x_t(S)|$ reaches $2c(n \log \log \log n)^{1/2}$ for some set $S$, we label $S$ dangerous and set its $\Delta_S = n^{1/2}/\log n$. This ensures that the discrepancy increment $\gamma_t(S)$ will have standard deviation at most $\gamma(n^{1/2}/\log n)$ henceforth, making $S$ extremely unlikely to incur an additional $cn^{1/2}$ discrepancy over the next $O(1/\gamma^2)$ steps. However, reducing the $\Delta_S$ comes at the price of increasing the entropy contribution of set $S$ in the left hand side of (3). Indeed, for the algorithm to be able to proceed, we need to ensure that (3) still holds with these reduced $\Delta_S$ (otherwise, we cannot guarantee the existence of the update vectors $\gamma_t$ with required properties).

To show that (3) still holds, we use two facts. First, that only a small fraction of sets will get dangerous. Second, the entropy contribution of each dangerous set is not too high. In particular, by Lemma 2.2, at most $2\exp(-2\log \log \log n) = 2(\log \log n)^{-2}$ fraction of sets ever get dangerous during the $1/\gamma^2$ steps. So, with probability at least $1/2$, the number of dangerous sets never exceeds $4n(\log \log n)^{-2}$. We condition on this event. On the other hand, each dangerous set $S$ contributes $g(\Delta_S/|S|^{1/2}) \le g(1/\log n) \le K \log \log n$ to (3), and hence the total entropy contribution of dangerous sets (conditioned on the event above) is $O(n/(\log \log n)^2) \cdot K \log \log n = o(n)$. Thus (3) will continue to hold, if there was some (reasonably small) slack to begin with.

A refinement of this idea, by considering multiple dangerous levels, allows us to reduce the discrepancy down to $O(n^{1/2})$ implying theorem 1.1.

## 4 An pseudo-approximation for Discrepancy

We prove theorem 1.3. Let $(V, \mathcal{S})$ be a set system, $V = [n]$, $\mathcal{S} = \{S_1, \ldots, S_m\}$ with hereditary discrepancy $\lambda$. For any $x \in \mathbb{R}^n$, let $x(S_j)$ denote the $\sum_{i \in S_j} x(i)$. Our algorithm will construct the final coloring iteratively in several steps. Let $x_t \in \mathbb{R}^n$ denote the coloring at time $t$. We start with $x_0 = (0, 0, \ldots, 0)$ initially. At each time step $t$, we update $x_t = x_{t-1} + \gamma_t$ for some suitably chosen vector $\gamma_t \in \mathbb{R}^n$. At the end, the final solution $x_f \in \{-1,+1\}^n$ will satisfy that $x_f(S_j) = O(\lambda \log(mn))$ for each $j \in [m]$.

During the algorithm, if element $i$ reaches $+1$ or $-1$ at time $t$, i.e. $x_t(i)$ becomes $+1$ or $-1$, we say that $i$ is *fixed* and it will never be updated again. A variable is *alive* at beginning of time $t$, if it has not been fixed by time $t - 1$. Let $A(t)$ denote the set of alive variables at end of time $t$. So, $A(0) = [n]$, and $A = \emptyset$ at the end, and moreover $|A(t)|$ is non-increasing with $t$. Let us assume that the algorithm knows $\lambda$ (it can try out all possible values for $\lambda$). We now describe the algorithm.

## 4.1 Algorithm

Initialize, $x_0(i) = 0$ for all $i \in [n]$. Let $s = 1/(4n(\log(mn))^{1/2})$. Let $\ell = 8 \log n/s^2$.
For each time step $t = 1, 2, \ldots, \ell$ repeat the following:

1. Find a feasible solution to the following semidefinite program:

$$\|\sum_{i \in S_j} v_i\|_2^2 \leq \lambda^2 \quad \text{for each set } S_j \tag{4}$$

$$\|v_i\|_2^2 = 1 \quad \forall i \in A(t-1) \tag{5}$$

$$\|v_i\|_2^2 = 0 \quad \forall i \notin A(t-1) \tag{6}$$

This SDP is feasible as setting $v_i \cdot v_j = \mathcal{X}(i)\mathcal{X}(j)$, where $\mathcal{X}$ is the minimum discrepancy coloring of the set system restricted to $A(t-1)$ is a valid solution. Let $v_i \in \mathbb{R}^n$, $i \in [n]$ denote some arbitrary feasible solution to the SDP above.

2. Construct $\gamma_t \in \mathbb{R}^n$ as follows: Let $g \in \mathbb{R}^n$ be obtained by choosing each coordinate $g(i)$ independently from the distribution $\mathcal{N}(0,1)$. For each $i \in [n]$, let $\gamma_t(i) = s\langle g, v_i \rangle$.
Update $x_t = x_{t-1} + \gamma_t$.
If $|x_t(i)| > 1$, for any $i$, abort the algorithm.

3. For each $i$, set $x_t(i) = 1$ if $x_t(i) \geq 1 - 1/n$ or set $x_t(i) = -1$ if $x_t(i) < -1 + 1/n$.
Update $A(t)$ accordingly.

Return the final coloring $x_\ell$.

## 4.2 Analysis

We begin with some simple observations.

1. At each time step $t$, we have $\|v_i\|_2^2 = 1$ for each $i \in A(t-1)$ and $\|v_i\|_0^2 = 0$ for $i \notin A(t-1)$. Thus, by lemma 2.1, conditioned on $i \in A(t-1)$, we have $\gamma_t(i) \sim N(0, s^2)$ for $i \in A(t-1)$ and $\gamma_t(i) = 0$ otherwise. Similarly, conditioned on the evolution of the algorithm until $t-1$, the increment $\gamma_t(S_j)$ for $S_j$ at time $t$ is an unbiased Gaussian with variance at most $s^2\lambda^2$ (the precise value of the variance will depend on $v(S_j) = \sum_{i \in S_j : i \in A(t-1)} v_i$, which depends on the SDP solution at time $t$, which in turn depends on the evolution of the algorithm until time $t-1$, in particular on the set of alive variables $A(t-1)$).

2. The rounding in step 3 of the algorithm can effect the overall discrepancy by at most $n \cdot (1/n) = 1$, as each variable is rounded up or down at most once and is never modified thereafter. Note $\lambda \geq 1$, unless the set system is empty, so we will ignore the effect of this rounding step henceforth.

3. For the algorithm to abort in step 2 at time $t$, it is necessary that $\gamma_t(i) > 1/n = 4s(\log n)^{1/2}$, as step 3 ensures that $|x_{t-1}(i)| < 1 - 1/n$. Since $\gamma_t(i)$ is distributed as $N(0, s^2)$, this probability is at most $\exp(-8 \ln mn) = (mn)^{-8}$. Since there at most $n$ variables and only $\ell = O(n^2 \log^2(mn))$ time steps, by union bound the probability that the algorithm ever aborts due to this step is at most $1/(mn)^4$.

The following key lemma shows that the number of alive variables halves in $O(1/s^2)$ steps with reasonable probability. The proof below follows a simpler presentation due to Joel Spencer.

**Lemma 4.1.** *Suppose $y \in [-1, +1]^n$ be an arbitrary coloring with at most $k$ alive variables. Let $z$ be the coloring obtained after applying steps (1)-(3) of our algorithm for $8/s^2$ time units. Then the probability that $z$ has $k/2$ or more alive variables is at most $1/4$.*

*Proof.* For $1 \leq t \leq u = 8/s^2$, let $y_t$ denote the coloring at time $t$ starting from $y$, i.e. after $t$ applications of steps (1)-(3). Let $K$ be the set of alive variables at $t = 0$. Let $k_t$ denote the number of variables alive the end of time $t$. For each time $t$, let us define $r_t = \sum_i y_t(i)^2$ if $k_{t-1} \geq k/2$. Otherwise, define $r_t = r_{t-1} + s^2 k/2$. Now, we claim that conditioned on any coloring $y_{t-1}$, the increment $r_t - r_{t-1}$ is at least $s^2 k/2$ in expectation (over the gaussian $g \in R^n$ at time $t$). This is clearly true if $k_t < k/2$. Otherwise if $k_t \geq k/2$, then

$$\begin{aligned} \mathbb{E}[r_t - r_{t-1}|y_{t-1}] &= \mathbb{E}[r_t|y_{t-1}] - r(t-1) \\ &= \mathbb{E}_g\left[\sum_i (y_{t-1}(i) + \gamma_t(i))^2\right] - \sum_i y_{t-1}(i)^2 \\ &= \sum_i \left(2y_{t-1}\mathbb{E}[\gamma_t(i)] + \mathbb{E}[\gamma_t(i)^2]\right) \geq s^2 k_{t-1} \geq s^2 k/2. \end{aligned}$$

The last step follows as $\mathbb{E}_g[\gamma_t(i)] = 0$ and $\mathbb{E}_g[\gamma_t(i)^2] = s^2$ for each alive variable in $y_{t-1}$ and is $0$ otherwise.

If there are still at least $k/2$ alive variables at $t = u$, then $r_u = \sum_{i \in K} y_t(i)^2 \leq k$. Moreover, for any run of the algorithm, it holds that $r_u \leq k + us^2 k/2$. This is because as long as $k_t \geq k/2$ it must be that $r_t \leq k$, but if $k_t$ becomes less than $k/2$, $r_t$ increases by exactly $s^2 k/2$ at each subsequent time step. Combining these facts we have,

$$us^2 k/2 \leq \mathbb{E}[r_u] \leq \Pr[k_u \geq k/2] \cdot k + (1 - \Pr[k_u \geq k/2]) \cdot (k + us^2 k/2)$$

and hence

$$\Pr[k_u \geq k/2] \leq \frac{k}{us^2 k/2} = 1/4.$$

□

Let $E$ denote the event that the final coloring $x_\ell$ is a proper $\{-1, +1\}$ coloring.

**Lemma 4.2.** $\Pr[E] \geq 1/n$. *That is, a proper coloring is produced with probability at least $1/n$.*

*Proof.* We apply lemma 4.1 with $y = x_t$ at epochs $t = 0, 8/s^2, 16/s^2, \ldots, (8 \log n)/s^2 = \ell$. As the number of alive variables initially is $n$, with probability at least $(1 - 1/4)^{\log n} \geq 1/n$, the number of alive variables reduces more than half at each epoch, and hence the number of alive variables is zero at $t = \ell$. □

We now prove theorem 1.3. Let $B_j$ denote the (bad) event that set $S_j$ has discrepancy more than $2 \log^{1/2}(mn) \cdot \lambda s \ell^{1/2}$ at the end of time step $\ell$. Let $B = B_1 \vee B_2 \vee \ldots \vee B_m$, and let $B^c$ denote the complement of $B$. To prove theorem 1.3, it suffices to show that $\Pr[B^c \cap E] \geq 1/(2n)$. Since $\Pr[B^c \cap E] \geq \Pr[E] - \Pr[B]$ and $\Pr[E] \geq 1/n$ by Lemma 4.2, it suffices to show that $\Pr[B] \leq 1/2n$.

As $x_t(S_j) = \sum_{t'=1}^{t} \gamma_{t'}(S_j)$ forms a martingale, with each increment $\gamma_t$ distributed (conditional upon the history until $t - 1$) as unbiased Gaussian with variance at most $\lambda^2 s^2$, by lemma 2.2 we have $\Pr[B_j] = \Pr[|x_\ell(S_j)| \geq 2 \log^{1/2}(mn) \cdot \lambda s \ell^{1/2}] \leq 2 \exp(-2 \log(mn)) = 2/(m^2 n^2)$. By union bound over the $m$ sets, $\Pr[B] \leq 2/(mn^2) \leq 1/(2n)$ which implies the result.

## 4.3 Constructive version of Srinivasan's result

We prove theorem 1.2. Let $n$ denote the number of elements, and let $m$ denote the number of sets. Since, each element lies in at most $t$ sets, we can assume that $m \leq nt$. The algorithm is essentially identical to that in section 4. The only difference is that, at any step $t$ in the algorithm, the entropy method, as applied in [19], only guarantees us a partial coloring (instead of a complete coloring) of the alive variables $A(t-1)$ with discrepancy $ct^{1/2}$. So we modify the first step of the algorithm above as follows:

Find a feasible solution to the following semidefinite program:

$$||\sum_{i \in S_j} v_i||_2^2 \leq c^2 t \qquad \text{for each set } S_j \tag{7}$$

$$\sum_{i \in A(t-1)} ||v_i||_2^2 \geq |A(t-1)|/2 \tag{8}$$

$$||v_i||_2^2 \leq 1 \qquad \forall i \in A(t-1) \tag{9}$$

$$||v_i||_2^2 = 0 \qquad \forall i \notin A(t-1) \tag{10}$$

The constant $c$ is not stated explicitly in [19], but it can be calculated (in fact our algorithm can do a binary search on $c$ do determine the smallest value $c$ for which the SDP has a feasible solution). This program is feasible, as $v_i(1) = \mathcal{X}(i)$, where $\mathcal{X}$ is the partial coloring of $A(t-1)$ with discrepancy $ct^{1/2}$, is a feasible solution.

The analysis is essentially identical to that in section 5. As in lemma 4.1, during $16/s^2$ steps, the number of alive variables reduces by a factor of 2, with probability at least $1/2$ (note that we have $16/s^2$ steps above instead of $8/s^2$ steps in Lemma 4.1, because of the partial coloring instead of complete coloring of $A(t-1)$). Thus, there is a proper coloring with probability at least $1/n$ at end of $(16/s^2) \cdot \log n$ steps. The expected discrepancy of each set $S$ in this coloring is at most $t^{1/2}(\log n)^{1/2}$. As there at most $nt$ sets, arguing as at the end of section 4.2, conditioned on obtaining a proper coloring at the end, each set has discrepancy at most $O((t \log n)^{1/2}(\log(nt))^{1/2}) = O(t^{1/2} \log n)$.

## 5 Constructive version of Spencer's result

In this section we prove theorem 1.1. In fact, we will prove the more general guarantee for $O(n^{1/2} \log(2m/n))$ for set systems with $n$ elements and $m$ sets, where $m \geq n$.

To show this, we will design an algorithmic subroutine with the following property.

**Theorem 5.1.** *Let $x \in [-1, 1]^n$ be some fractional coloring with at most $a$ alive variables (i.e. $i$ with $x(i) \notin \{-1, +1\}$). Then, there is an algorithm that with probability at least $1/2$, produces a fractional coloring $y \in [-1, 1]^n$ with at most $a/2$ alive variables, and the discrepancy of any set increases by at most $O(a^{1/2} \log(2m/a))$.*

Given theorem 5.1, the main result follows easily.

**Lemma 5.2.** *The procedure in theorem 5.1 implies an algorithm to find a proper $\{-1, +1\}$ coloring with discrepancy $O(n^{1/2} \log(2m/n))$. Moreover, the algorithm succeeds with probability at least $1/(2 \log m)$.*

*Proof.* We start with the coloring $x = (0, 0, \ldots, 0)$, and apply theorem 5.1 for $\ell = \log \log m$ steps. With probability at least $2^{-\ell} = 1/\log m$, this gives a fractional coloring $y$ with at most $n/2^\ell = n/\log m$ alive variables, with the property that the discrepancy $y(S)$ of any set is at most

$$\sum_{k=1}^{\ell} O\left(\left(\frac{n}{2^k}\right)^{1/2} \log\left(\frac{m 2^{k+1}}{n}\right)\right) = O\left(n^{1/2} \log\left(\frac{2m}{n}\right)\right).$$

Finally, to obtain a proper coloring $z$ from $y$, we randomly round each alive variable $i$, i.e. set $z(i) = -1$ with probability $(1 - y(i))/2$ or to $+1$ with probability $(1 + y(i))/2$.

In expectation, $\mathbb{E}[z(i)] = y(i)$. Since there at most $n/\log m$ variables, by Chernoff bounds, the probability that a set $S$ incurs an additional discrepancy of $c(n/\log m)^{1/2}$ is at most $2e^{-c^2/2}$. Thus, choosing $c = 2\log^{1/2} m$, with high probability every set incurs an additional discrepancy of $O(n^{1/2}) \leq O(n^{1/2} \log(2m/n))$. □

We will focus on proving theorem 5.1 henceforth. We first describe the subroutine, and then analyze it.

## 5.1 Algorithmic Subroutine

Consider the following subroutine. The input is a coloring $x_0 \in [-1, +1]^n$ with at most $a$ alive variables. Let $s = 1/(4\log^{3/2}(mn))$, and let $q = \log(2m/a)$. Let $d = 9\log(20K)$ and let $c = 64(d(1+\ln K))^{1/2}$ be constants where $K$ is defined as in (3). For each time $t = 1, 2, \ldots$ repeat the following steps until $t = 16/s^2$ or fewer than $a/2$ variables are alive, whichever occurs earlier.

1. For each set $S_j$, let $\eta_j$ denote the total discrepancy incurred by $S_j$ thus far, i.e. $\eta_j = \left|\sum_{s=1}^{t-1} \gamma_s(S_j)\right|$. Define $\beta(0) = 0$ and for $k = 1, 2, \ldots,$ define

$$\beta(k) = ca^{1/2}(q+1)\left(2 - \frac{1}{k}\right).$$

    For $k = 0, 1, 2, \ldots,$ we say that $S_j$ is $k$-dangerous at time $t$ if $\eta_j \in [\beta(k), \beta(k+1))$.

    If $\eta_j > 2\beta(1)$ ( note that $2\beta(1) \geq \beta(k)$ for any $k$) for any $j$, abort the algorithm and return fail.

2. For $k = 0, 1, 2 \ldots,$ let $\mathcal{S}(k) \subseteq \mathcal{S}$ denote the sub-collection of sets that are currently $k$-dangerous. Let $A(t-1)$ denote the set of variables that are currently alive. For $k = 0, 1, \ldots,$ define

$$\alpha(k) = \frac{da(q+1)}{(k+1)^5}.$$

    Find a feasible solution to the following semidefinite program:

$$\sum_{i \in [n]} \|v_i\|_2^2 \geq A(t-1)/2 \tag{11}$$

$$\left\|\sum_{i \in S_j} v_i\right\|_2^2 \leq \alpha(k) \qquad \forall k = 0, 1, 2, \ldots, \; \forall S_j \in \mathcal{S}(k) \tag{12}$$

$$\|v_i\|_2^2 \leq 1 \qquad \forall i \in A(t-1) \tag{13}$$

$$\|v_i\|_2^2 = 0 \qquad \forall i \notin A(t-1) \tag{14}$$

    If the SDP does not have feasible solution, abort the algorithm and return fail.
    Otherwise, let $v_i \in \mathbb{R}^n$, $i = 1, \ldots, n$ be the solution returned by the SDP.

3. We construct $\gamma_t$ from these $v_i$ as follows: Let $g \in \mathbb{R}^n$ be obtained by choosing each coordinate $g(i)$ independently $\mathcal{N}(0,1)$. For each $i \in [i]$, let $\gamma_t(i) = s\langle g, v_i \rangle$. Update $x_t = x_{t-1} + \gamma_t$. Abort the algorithm if $|x_t(i)| > 1$ for any $i$.

4. For each $i$, if $x_t(i) \geq 1 - 1/\log(mn)$, set $x_t(i) = 1$ with probability $(1 + x_t(i))/2$ or to $-1$ otherwise. Similarly, if $x_t(i) < -1 + 1/\log(mn)$, set $x_t(i) = -1$ with probability $(1 - x_t(i))/2$ or to $+1$ otherwise. Update $A(t)$ accordingly.

## 5.2 Analysis

We first note some simple observations.

1. For the algorithm to abort in step 3, it must be the case that $\gamma_t(i) > 1/\log(mn)$ for some $t, i$ (this is ensured by step 4 of the algorithm). However, since $s = 1/(4\log^{3/2}(mn))$, this happens with probability at most $1/(m^4 n^4)$ and hence we ignore its effect henceforth.

2. The rounding in step 4 adds an overall discrepancy of $O(a^{1/2})$ to every set, during the course of the subroutine. This is because, the variance incurred when a variable is rounded in step 4 is $O(1/\log(mn))$. Since at most $a$ variables will ever be rounded, the variance for any constraint is $O(a/\log mn)$. The result now follows by standard tail bounds and taking union over the $m$ sets.

The following lemma gives a sufficient condition for the SDP to be feasible.

**Lemma 5.3.** *Consider any time $t$. If for every $k = 1, 2, \ldots$ no more than $m_k = a2^{-10(k+1)}/K$ sets are $k$-dangerous at $t$, then the SDP defined by (11)-(14) has a feasible solution.*

*Proof.* We will show that if the conditions of the lemma hold, then by the entropy method, there exists a feasible partial coloring $\mathcal{X}$ on at least $|A(t-1)|/2$ elements such that $|\mathcal{X}(S_j)| \leq \Delta_{S_j} = (\alpha(k))^{1/2}$ is satisfied for each $k$-dangerous set $S_j$, for $k = 0, 1, 2, \ldots$. As $\mathcal{X}$ gives a feasible solution to the SDP constraints (11)-(14), this will imply the result.

Thus, it suffices to show that condition (3) holds for the given choice of $m_k$ and $\Delta_{S_j}$. That is,

$$\sum_{j \in [m]} g(\lambda_j) \leq \frac{1}{5}(a/2) \leq \frac{1}{5}|A(t-1)| \qquad (15)$$

where $\lambda_j = \Delta_{S_j} \cdot (|S_j \cap A(t-1)|)^{-1/2}$. Since $g(\lambda)$ is a decreasing function of $\lambda$, to prove (15), we can use any lower bound on $\lambda_j$. For any $k$-dangerous set $S_j$, for $k = 0, 1, \ldots$,

$$\lambda_j = \Delta_{S_j} \cdot (|S_j \cap A(t-1)|)^{-1/2} \geq (\alpha(k))^{1/2}(|A(t-1)|)^{-1/2} \geq (d(q+1)(k+1)^{-5})^{1/2}.$$

Let us define $\zeta(k) = (d(q+1)(k+1)^{-5})^{1/2}$.

We now upper bound the left hand side of (15). As $\zeta(0) = (d(q+1))^{1/2} \geq 0.1$, the contribution of $0$-dangerous sets to the left hand side of (15) is at most

$$m \cdot K \cdot \exp(-\zeta(0)^2/9) = m \cdot K \cdot \exp(-d(q+1)/9) \leq \frac{1}{20}m \exp(-q-1) \leq \frac{a}{20}. \qquad (16)$$

We now bound $\sum_{k \geq 1} m_k \cdot g(\zeta(k))$. For any $k \geq 1$, we have

$$g(\zeta(k)) \leq K \cdot \max(\ln(10), \ln(1/\zeta(k))) \leq K \cdot \max(\ln(10), \ln((k+1)^{5/2})) \leq 5K \ln(k+1).$$

Thus,

$$\sum_{k \geq 1} m_k \cdot g(\zeta(k)) \leq \sum_{k \geq 1} \frac{1}{K} a 2^{-10(k+1)} \cdot 5K \ln(k+1) \leq a/20. \qquad (17)$$

By (16) and (17) it follows that (15) holds, which proves the lemma. □

**Lemma 5.4.** *For $k = 1, 2, \ldots$, let $D_k$ denote the event that more than $m_k = a2^{-10(k+1)}/K$ sets ever become $k$-dangerous during $t = 1, \ldots, 16/s^2$. It holds that $\Pr[D_k] \leq 2^{-5(k+1)}$.*

*Proof.* We first prove the claim for $k = 1$. Suppose some set $S_j$ becomes 1-dangerous at some time. Then, there must be a time $\hat{t}$ when $|\eta_j|$ first exceeds $\beta(1)$. However, until $\hat{t}$, $\eta_j$ was evolving as martingale, with each conditional increment distributed as an unbiased Gaussian with variance at most $\alpha(0)s^2$. By lemma 2.2, this has probability at most

$$2\exp\left(-\frac{\beta(1)^2}{2\alpha(0)s^2(16/s^2)}\right) \leq \exp\left(-\frac{c^2(q+1)}{64d}\right)$$
$$= \exp(-64(q+1)(1+\ln(K))) \leq \frac{1}{K}2^{-60}2^{-q-1} = \frac{1}{K}2^{-60}\frac{a}{m}. \quad (18)$$

Thus the expected number of such sets is at most $a(1/K)2^{-60}$ and hence the claim for $k = 1$ holds by Markov's inequality.

For $k \geq 2$, the argument is similar. For $S_j$ to become $k$-dangerous during phase $q$, it must have become $k-1$-dangerous at some time $\hat{t}$ during phase $q$ and then traversed the distance $\beta(k) - \beta(k-1)$ during at most $16/s^2$ time steps[1]. Since $\gamma_t(S_j)$ (the conditional increment of $\eta_j$) has variance most $\alpha(k-1)s^2$ whenever $\eta_j \in [\beta(k-1), \beta(k)]$, due to the SDP constraint (12), Lemma 2.2 implies that the probability that $S_j$ becomes $k$-dangerous at any time is at most

$$\exp\left(-(\beta(k) - \beta(k-1))^2/(4\alpha(k-1)s^2 \cdot (16/s^2))\right) \leq \exp\left(-(c^2(q+1)k)/(64d)\right)$$
$$= \exp(-64(q+1)(1+\ln K)k) \leq \frac{1}{K} \cdot 2^{-q-1} \cdot 2^{-32(k+1)}$$

By Markov's inequality, $\Pr[D_k] \leq 2^{-5(k+1)}$, which proves the lemma. □

We can now finish off the proof of theorem 5.1. Let $D = \vee_{k=1}^\infty D_k$, and let $E$ denote the event that the number of alive variables is more than $a/2$ at $t = u = 16/s^2$. Let $D^c$ and $E^c$ denote the complement of $D$ and $E$. Note that if $D^c$ holds, then by Lemma 5.3, the SDP is always feasible, and the algorithm never aborts in step 2 of the algorithm. Moreover, as $m_k \ll 1$ for $k = c(\log m)$ for large enough $c$, it follows that if $D_k^c$ holds then no set ever incurs a discrepancy of more than $\beta(k) \leq 2\beta(1)$.

Now to prove theorem 5.1 it suffices to show that $\Pr[D^c|E^c] \geq 1/2$.

By Lemma 5.4, $\Pr[D] \leq \sum_{k \geq 1} \Pr[D_k] \leq 1/16$. Also, $\Pr[E] \leq 1/4$ follows by an argument identical to that in the proof of lemma 4.1. In particular, if the number of alive variables at $t$ is at least $a/2$, we set $r_t = \sum_i x_t(i)^2$, otherwise, we set $r_t = r_{t-1} + s^2a/4$. Thus, irrespective of $x_{t-1}$, the increment $r_t - r_{t-1}$ increases in expectation by

$$\sum_i \gamma_t(i)^2 = \sum_{i \in A(t-1)} s^2 \|v_i\|_2^2 \geq s^2a/4.$$

Moreover, as $r_t$ can never exceed $a + ts^2a/4$, it follows that after $u$ steps,

$$us^2a/4 \leq \mathbb{E}[r_t] \leq \Pr(E) \cdot a + (1 - \Pr(E)) \cdot (a + us^2a/4)$$

implying that $\Pr[E] \leq 4/(us^2) = 1/4$.

Thus, $\Pr[D^c|E^c] \geq \Pr[D^c \cap E^c] \geq 1 - \Pr[D] - \Pr[E] \geq 1/2$, and the result follows.

---

[1]Strictly speaking, there is a non-zero probability that a $k - 2$ or less dangerous set may become $k$-dangerous at next step, however this probability is super-polynomially small as $(\beta(k+1) - \beta(k))/s^2\alpha(k) \geq \log^2 n$ (and $\alpha(k) \approx \alpha(k-1)$). Moreover, it can be made arbitrarily small by setting $s$ arbitrarily small, say $1/n$. So, we can ignore this event in the analysis.

## Acknowledgments

We thank Joel Spencer for several insightful comments that lead to a simpler and clearer presentation of the results in this paper. Also, in an earlier version of the paper we had erroneously claimed that the techniques in theorem 1.3 can be used to approximate the hereditary discrepancy within a logarithmic factor. We thank the anonymous FOCS referees for pointing out this error to us. Finally, we thank Konstantin Makarychev, Jiri Matousek, Viswanath Nagarajan, Aravind Srinivasan and Ola Svensson for several useful comments about the paper and discrepancy in general.